\documentclass[aps,preprintnumbers,amsmath,amssymb,nofootinbib,superscriptaddress]{revtex4}
\usepackage{amsfonts}
\usepackage{amsmath}
\usepackage{amssymb}
\usepackage{array}
\usepackage{bm}
\usepackage{bbm}
\usepackage{color}
\usepackage{epsfig}
\usepackage{float}
\usepackage{graphicx}
\usepackage{hyperref}
\usepackage{lipsum}
\usepackage{mathrsfs}
\usepackage{multirow}
\usepackage{ulem}
\usepackage{upgreek}

\allowdisplaybreaks

\makeatletter
\def\hlinew#1{%
  \noalign{\ifnum0=`}\fi\hrule \@height #1 \futurelet
   \reserved@a\@xhline}
\usepackage{array}
\newcommand{\PreserveBackslash}[1]{\let\temp=\\#1\let\\=\temp}
\newcolumntype{C}[1]{>{\PreserveBackslash\centering}p{#1}}
\newcolumntype{R}[1]{>{\PreserveBackslash\raggedleft}p{#1}}
\newcolumntype{L}[1]{>{\PreserveBackslash\raggedright}p{#1}}

\newcommand{\jpsi}{{J/\psi}}

\newcommand{\nn}{\nonumber}
\newcommand{\beq}{\begin{equation}}
\newcommand{\eeq}{\end{equation}}
\newcommand{\bqa}{\begin{eqnarray}}
\newcommand{\eqa}{\end{eqnarray}}
\newcommand{\tabincell}[2]{\begin{tabular}{@{}#1@{}}#2\end{tabular}}

\begin{document}

\preprint{JLAB-THY-24-4164}

\title{\mbox{}\\[10pt]
Two-loop QCD corrections to Higgs radiative decay to vector quarkonium
}

\author{Yu Jia\footnote{jiay@ihep.ac.cn}}
\affiliation{Institute of High Energy Physics, Chinese Academy of
Sciences, Beijing 100049, China\vspace{0.2cm}}
\affiliation{School of Physical Sciences, University of Chinese Academy of Sciences,
Beijing 100049, China\vspace{0.2cm}}

\author{Zhewen Mo\footnote{mozw@itp.ac.cn}}
\affiliation{CAS Key Laboratory of Theoretical Physics, Institute of Theoretical Physics, Chinese Academy of Sciences,
Beijing 100190, China\vspace{0.2cm}}
\affiliation{Institute of High Energy Physics, Chinese Academy of
Sciences, Beijing 100049, China\vspace{0.2cm}}

\author{Jia-Yue Zhang\footnote{jzhang@jlab.org}}
\affiliation{Theory Center, Jefferson Lab, Newport News, Virginia 23606, USA}
\affiliation{Institute of High Energy Physics, Chinese Academy of
Sciences, Beijing 100049, China\vspace{0.2cm}}

\date{\today}

\begin{abstract}
The exclusive production of $J/\psi$ through Higgs boson radiative decay may serve a clean channel to extracting the charm quark Yukawa coupling. We calculate the two-loop QCD corrections to $H\rightarrow J/\psi(\Upsilon)+\gamma$ using an optimized nonrelativistic QCD (NRQCD) approach. We compute the ${\cal O}(\alpha_s^2)$ correction in the direct channel, where Higgs directly couples to $c\bar{c}$, as well as the ${\cal O}(\alpha_s)$ correction in the indirect channel, {\it viz.}, $H\to\gamma^*\gamma$ followed by the virtual photon fragmentation into $J/\psi$. Incorporating the destructive interference between the direct and indirect channels, we present to date the most accurate predictions for Higgs boson radiative decay into vector quarkonium, $\mathcal{B}(H\rightarrow J/\psi+\gamma) = 3.27_{-0.07}^{+0.30}{}^{+0.06}_{-0.06}{}_{-0.13}^{+0.13}\times 10^{-6}$, and $\mathcal{B}(H\rightarrow \Upsilon+\gamma)= 1.34_{-0.31}^{+0.75}{}^{+0.25}_{-0.20}{}^{+0.05}_{-0.05}\times 10^{-8}$.
\end{abstract}

\maketitle

\section{Introduction}

The ground-breaking discovery of a $125$ GeV boson by the {\tt ATLAS}~\cite{ATLAS:2012yve} and {\tt CMS}~\cite{CMS:2012qbp} collaboration in $2012$
marks a milestone in the history of particle physics. After intensive investigations over a decade,
this new boson has been firmly identified with the long-sought Higgs boson in Standard Model (SM).
While the measured Higgs and gauge boson couplings are in perfect agreement with what is expected from the Higgs mechanism,
the experimental test of the Higgs Yukawa couplings is still incomplete.
To date, the experimental constraints on the Yukawa couplings are mainly
involving the third-generation quarks~\cite{ATLAS:2016neq}.
In particular, both the sign and the magnitude of the Higgs-charm Yukawa coupling still remains elusive.

Since the Yukawa coupling is proportional to fermion mass, it is rather challenging for \texttt{LHC} to pin down the
Yukawa couplings of the first two generations of quarks.
The signal events of Higgs boson decays into $c\bar{c}$ at {\tt LHC} are overwhelmed by the copious QCD background,
meanwhile it is also challenging to distinguish jets with different flavors.
Nevertheless, it is anticipated that the high-luminosity LHC ({\tt HL-LHC})~\cite{ATLAS:2013qma,CMS:2013aga,Dawson:2013bba}
and prospective $e^+e^-$ colliders such as {\tt FCC} and {\tt CEPC} may have a good chance to measure the charm Yukawa coupling.

It has been proposed that the charm Yukawa coupling may be measured at \texttt{HL-LHC} through
Higgs boson production in association with a charm-tagged jet, {\it i.e.},
$pp\to H+c$~\cite{Brivio:2015fxa}, Higgs decay to $c\bar{c}+\gamma$ by identifying the charm jets in the final states~\cite{Han:2017yhy,Han:2018juw, Carlson:2021tes},
the decay of Higgs to a vector charmonium and a gauge boson ($\gamma, Z$)~\cite{Bodwin:2013gca,Kagan:2014ila},
and Higgs boson inclusive decay to charmonia via $c$-quark fragmentation~\cite{Han:2022rwq}.
To date, the most stringent constraint on charm Yukawa coupling, $1.1<|\kappa_c|<5.5$,
is placed by the {\tt CMS} Collaboration by analyzing the $c\bar{c}$ jets in the final state,
with the aid of machine learning techniques~\cite{CMS:2022psv}.

The rare exclusive decay channel $H\to J/\psi+\gamma$ looks attractive for measuring the charm Yukawa coupling,
since the $J/\psi$ can be cleanly reconstructed through leptonic decay at hadron colliders.
The \texttt{ATLAS} collaboration recently searched several channels of
Higgs radiative decay into vector quarkonia~\cite{ATLAS:2022rej}.
The upper limits of the branching ratios of Higgs radiative decays into $\jpsi$, $\psi(2S)$ and $\Upsilon(1S,2S,3S)$ have been placed,
which are $2.1\times 10^{-4}$, $10.9\times 10^{-4}$, and $(2.6, 4.4, 3.5)\times 10^{-4}$, respectively~\cite{ATLAS:2022rej}.

It is worth pointing out that, the $\jpsi$ production in Higgs radiative decay is mediated by two distinct mechanisms.
In the {\it direct} channel, Higgs boson first radiative decays into a comoving $c\bar{c}$ pair in short distance,
which subsequently evolves into the $J/\psi$ state after emitting a hard photon.
On the other hand, in the {\it indirect} channel, the Higgs boson first disintegrates into two photons via the $W/t$ loop,
followed by one virtual photons fragmentating into $J/\psi$.
For $H\to J/\psi\gamma$ process, the indirect amplitude, which by itself is insensitive to the charm Yukawa coupling,
is however actually about 5 times greater than the direct amplitude in magnitude.
In $2013$, Bodwin {\it et al.}~\cite{Bodwin:2013gca}, has made a comprehensive lowest-order investigation on $H\to J/\psi\gamma$,
and found that the direct and indirect amplitudes interfere destructively.
Bodwin {\it et al.}~\cite{Bodwin:2013gca} pointed out that it would be
hopeless to measure the charm Yukawa coupling at {\tt LHC} if only retaining the direct channel contribution. Nevertheless,
thanks to the destructive interference effect, the decay rate of this Higgs rare decay process would be considerably enhanced  so that
the charm Yukawa coupling might be accessible at {\tt LHC}.

In contrast, the direct and indirect amplitudes for $H\to \Upsilon\gamma$, accidentally, have comparable magnitude yet opposite sign, so that
the destructive interference would lead to an exceedingly tiny branching fraction.
It is unlikely to observe this process at {\tt LHC} within SM in foreseeable future.
Nevertheless, it has been proposed that $H\to \Upsilon\gamma$ would be a useful portal to probe some beyond-SM scenarios, {\it viz.},
the wrong sign $hb\bar{b}$ coupling~\cite{Modak:2016cdm,Batra:2022wsd} and $H\gamma\gamma$ anomalous coupling~\cite{Dong:2022bkd}.

Prior to the Higgs boson discovery, there are only scarce studies of $H\to \jpsi+\gamma$~\cite{Shifman:1980dk,Jia:2008ep},
which mainly concentrated on the direct channel. At lowest order in charm quark velocity,
the $\mathcal{O}(\alpha_s)$ correction to $H\to J/\psi+\gamma$ in the direct channel
was first calculated by Shifman and Vysotsky in $1980$~\cite{Shifman:1980dk}.
The NRQCD short-distance coefficients (SDCs) in this process contains large collinear logarithms of type
$\alpha_s^n \ln^n m_H^2/m_c^2$ at each perturbative order, which may potentially spoils perturbative convergence.
The resummation of the leading logarithm was first made by \cite{Shifman:1980dk}, and later recast in a modern effective-field-theory context,
by refactorizing the $\jpsi$ light-cone distribution amplitude (LCDA) onto NRQCD and then employing the
Efremov-Radyuskin-Brodsky-Lepage (ERBL) evolution equation to resum leading collinear logarithms~\cite{Jia:2008ep}.

In the post-Higgs discovery era, especially after the work by Bodwin {\it et al.} in 2013~\cite{Bodwin:2013gca},
an array of theoretical efforts have been invested on the $J/\psi$ production through Higgs radiative decay.
The main theoretical tools are the model-independent NRQCD factorization~\cite{Bodwin:1994jh} and
collinear factorization (light-cone) approaches, or the joint use of two approaches.
Bodwin {\it et al.} investigate the order-$v^2$ relativistic correction and summed the collinear logarithms
through relative order $\alpha_s^2$~\cite{Bodwin:2014bpa}. By assuming a model parametrization for the LCDA of $\jpsi$,
K\"{o}nig and Neubert predicted the branching fractions of $H\to \jpsi+\gamma$ within the collinear factorization, and resummed
the collinear logarithms through next-to-leading-logarithmic (NLL) accuracy~\cite{Konig:2015qat}.
Later Bodwin {\it et al.} have developed a new numerical algorithm, which is based on Abel summation and Pad\'e approximation,
to accomplish the NLL resummation for the NRQCD SDC~\cite{Bodwin:2016edd,Bodwin:2017wdu}.
In 2019 Brambilla {\it et al.} have improved the prediction of the direct amplitude by including the order-$v^4$
relativistic correction~\cite{Brambilla:2019fmu}.

The one-loop QCD correction to $H\to \jpsi\gamma$ in the direct channel is known to be negative and sizable~\cite{Shifman:1980dk,Zhou:2016sot},
which may reach $-50\%$ of the tree level contribution. It is naturally to speculate what is the magnitude of
even higher-order radiative corrections?
Furthermore, since the indirect channel makes a dominant contribution, it seems compulsory to make a more precise prediction
also for this leading channel, {\it e.g.}, by including the NLO QCD correction.
It is the very goal of this work to calculate the two-loop QCD corrections for both direct and indirect channels within
the NRQCD factorization framework,  and consequently provide the most precise predictions to $H\to \jpsi(\Upsilon)+\gamma$.

The rest of the paper is organized as follows.
In Sec.~\ref{Sec:nrqcd:fac} we introduce the optimized NRQCD factorization approach for Higgs boson radiative decay into vector quarkonium,
specifying the separate treatment of the direct and indirect channels.
In Sec.~\ref{Outline:Calculation}, we sketch the technical strategy of calculating the two-loop QCD corrections for $H\to \jpsi\gamma$.
In Sec.~\ref{SDCs:through:two:loop} we report the results of the SDCs in both direct and indirect channels through two-loop order in the
optimized NRQCD approach. While the one-loop expressions have been presented in analytic form, the two-loop SDCs are provided in
purely numerical format.
We devote Sec.~\ref{phenomenology} to a detailed phenomenological analysis, which presents
the finest predictions for the branching fractions of $H\to\jpsi (\Upsilon)+\gamma$.
Finally we summarize in Sec.~\ref{summary}.

\section{NRQCD factorization for Higgs radiative decay to vector quarkonium}
\label{Sec:nrqcd:fac}

To be definite, we will specify the vector quarkonium state from the Higgs radiative decay to be $J/\psi$.
Generalization to other vector quarkonia such as $\psi'$, $\Upsilon(nS)$ is straightforward.
Let $Q$, $P$, and $k$ represent the momenta of the Higgs boson, $J/\psi$, and $\gamma$, respectively.
By Lorentz and QED gauge invariance, the corresponding amplitude can be cast into the following form:
\beq
\mathcal{M}(H(Q)\rightarrow \jpsi(P)+\gamma(k))=\left[2k\cdot\varepsilon^*_\jpsi P\cdot\varepsilon^*_\gamma-
(m_H^2-m_\jpsi^2)\varepsilon^*_\jpsi\cdot\varepsilon^*_\gamma\right]\, F(\tau),
\eeq
where $\varepsilon_\jpsi$ and $\varepsilon_\gamma$ signify the polarization vectors of $J/\psi$ and $\gamma$, respectively.
All the nonperturbative dynamics is encapsulated in a single form factor $F(\tau)$ with $\tau\equiv 4m_c^2/m_H^2$,
which can be divided into two pieces:
\begin{align}
  F(\tau) =  F_\mathrm{dir}(\tau) + F_\mathrm{indir}(\tau),
\label{form:factor:divide:two:parts}
\end{align}
characterizing the contributions from the direct and indirect $J/\psi$ production mechanisms, respectively.
The partial width of $H\to \jpsi+\gamma$ then reads
\beq
\Gamma(H\to J/\psi+\gamma)= \frac{\left(m_H^2-m_\jpsi^2\right)^3}{8\pi m_H^3}
 \left|F_\text{dir}\left(\tau\right)+F_\text{indir}\left(\tau\right)\right|^2,
\label{partial:wdith:formula:H:to:jpsi:gamma}
\eeq

The hard exclusive quarkonium production process is amenable to NRQCD factorization at the amplitude level.
As will be elaborated later,  it turns out to be advantageous to employ the {\it optimized} NRQCD factorization approach,
where the direct and indirect $J/\psi$ production channels, being separately gauge-invariant, are treated differently.

\begin{figure}[hbt]
\begin{center}
\includegraphics[width=0.24\textwidth]{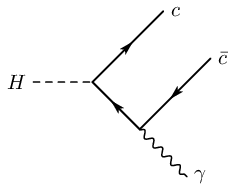}
\includegraphics[width=0.24\textwidth]{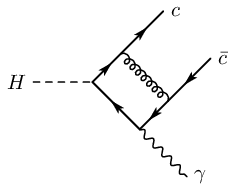}
\includegraphics[width=0.24\textwidth]{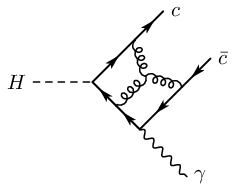}
\caption{Representative Feynman diagrams for the direct channel of $H\rightarrow c\bar{c}({}^3S_1^{(1)})+\gamma$ through ${\cal O}(\alpha_s^2)$.
\label{Feyn:Diagrams:Dir}}
\end{center}
\end{figure}

\subsection{Direct $J/\psi$ production mechanism}

As pictured in Fig.~\ref{Feyn:Diagrams:Dir}, in the direct channel an energetic color-singlet
$c\bar{c}$ pair is first created from Higgs decay,
then hadronizes into the $J/\psi$ after emitting a hard photon. At the lowest order (LO) in charm quark velocity $v$,
the form factor in the direct channel can be factorized into
\beq
F_\mathrm{dir}(\tau)=\sqrt{\dfrac{G_F}{\sqrt{2}}}  \left( \dfrac{4e e_c m_c}{m_H^2}\right)  {\mathcal{F}_\mathrm{dir}}(\tau,\mu_\Lambda)
    \dfrac{\langle \jpsi(\boldsymbol\epsilon) \vert \psi^\dagger \boldsymbol{\sigma}\cdot\boldsymbol{\epsilon}\chi(\mu_\Lambda) \vert 0 \rangle}{\sqrt{2m_c}}+{\mathcal O}(v^2),
\label{NRQCD:factorization:direct}
\eeq
with $G_F$ denoting the Fermi constant and $m_c$ representing the charm quark mass.
$\langle \jpsi(\boldsymbol\epsilon) \vert \psi^\dagger \boldsymbol{\sigma}\cdot\boldsymbol{\epsilon}\chi(\mu_\Lambda) \vert 0 \rangle$ signifying
the nonperturbative yet universal vacuum-to-$J/\psi$ LO NRQCD matrix element, and ${\mathcal{F}_\mathrm{dir}}(\tau,\mu_\Lambda)$ is the corresponding
short-distance coefficient (SDC). Although both the NRQCD matrix element and the respective SDC depend on the
NRQCD factorization scale $\mu_\Lambda$ logarithmically,
their product, or equivalently, the form factor $F_\mathrm{dir}$, should be independent of this artificial scale.

Since $m_H,m_c\gg \Lambda_\text{QCD}$, the asymptotic freedom of QCD allows the NRQCD SDCs to be computed
in perturbation theory, order by order in $\alpha_s$.  Through $O(\alpha^2_s)$, the SDC in the direct channel can be organized in
terms of perturbative order:
\beq
{\mathcal{F}}_\text{dir}
={\mathcal{F}}_\text{dir}^{(0)}+{\mathcal{F}}_\text{dir}^{(1)}+{\mathcal{F}}_\text{dir}^{(2)}+\cdots,
\label{form:factor:direct:decomposition}
\eeq
where the superscript signifies the powers of $\alpha_s$.
The one-loop and two-loop QCD corrections to the SDC in the direct channel can be parameterized as
\begin{subequations}
\bqa
 && {\mathcal{F}}_\text{dir}^{(1)}(\tau, \mu_R)
= {\mathcal{F}}_\text{dir}^{(0)}(\tau)  {\alpha_s(\mu_R)\over \pi} f^{(1)}(\tau),
\label{NRQCD:SDC:dir:one:loop}
\\
 && {\mathcal{F}}_\text{dir}^{(2)}(\tau, \mu_R,\mu_\Lambda)=
{\mathcal{F}}_\text{dir}^{(0)}(\tau) {\alpha_s(\mu_R)^2\over \pi^2} \bigg[ {\beta_0\over 4} \ln\! {\dfrac{4\mu_R^2}{m_H^2}} f^{(1)}(\tau)
+\gamma_\jpsi\ln\dfrac{\mu^2_{\Lambda}}{m_c^2} + f^{(2)}(\tau) \bigg],
\label{NRQCD:SDC:dir:two:loop}
\eqa
\label{NRQCD:SDC:dir:through:two:loop}
\end{subequations}
where the $C_F={4/3}$ and $C_A=3$ denote the Casimir of the $SU(3)$ color group, and
$\mu_R$ represents the renormalization scale.
$\beta_0=11 C_A/3-2n_f/3$ denotes the one-loop coefficient of the QCD $\beta$-function coefficient, with
$n_f=5$ denotes the number of active quark flavors relevant to Higgs decay.
Note that the explicit occurrences of the $\mu_R$ and $\mu_\Lambda$ first enter into two-loop SDC.
The occurrence of $\beta_0\ln \mu_R$ is dictated by the renormalization group invariance of ${\mathcal{F}}_\text{dir}$.
$\gamma_{\jpsi}=-{\pi^2}C_F(2C_F+3C_A)/{12}$ is the leading-order anomalous dimension of the NRQCD vector current $\psi^\dagger \boldsymbol{\sigma}\chi$,
which first arises at two-loop order. The occurrence of $\gamma_\jpsi \ln \mu_\Lambda$ at two-loop order is
demanded by the NRQCD factorization for exclusive process entailing $J/\psi$~\cite{Czarnecki:1997vz,Beneke:1997jm}.
The dimensionless functions $f^{(1,2)}(\tau)$ are affiliated with the one- and two-loop non-logarithmic contributions to the SDC.
One of the main tasks in this work is to compute the coefficient function $f^{(2)}(\tau)$.

\subsection{Indirect $J/\psi$ production mechanism}

\begin{figure}[hbt]
\begin{center}
\includegraphics[width=0.24\textwidth]{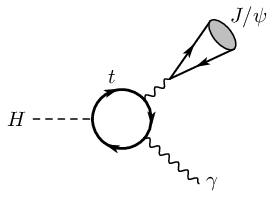}
\includegraphics[width=0.24\textwidth]{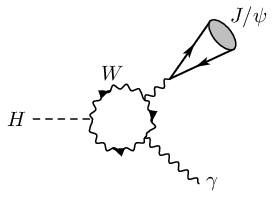}
\includegraphics[width=0.24\textwidth]{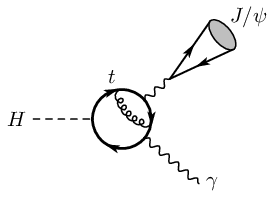}
\caption{Representative Feynman diagrams for the indirect channel of $H\rightarrow J/\psi+\gamma$ through ${\cal O}(\alpha_s)$.
\label{Feyn:Diagrams:Indir}}
\end{center}
\end{figure}

An alternative, numerically more significant, $\jpsi$ production mechanism from Higgs radiative decay is via the indirect channel, where
the Higgs boson first disintegrates into a real photon and a virtual photon,
followed by the virtual photons fragmentating into $J/\psi$. At SM, the $H\gamma\gamma$ coupling is absent in tree level.
However, the $H\to\gamma^*\gamma$ can be mediated through the top quark and $W$ boson loop.
Some representative diagrams are depicted in Fig.~\ref{Feyn:Diagrams:Indir}. Obviously such sub-processes have nothing to do with
the charm quark Yukawa coupling.
The corresponding form factor in the indirect channel reads
\beq
F_\mathrm{indir}(\tau) = \sqrt{G_F}\, {\mathcal{F}_\mathrm{indir}}(\tau)\left(\frac{-i}{m_\jpsi^2}\right) (i e e_c f_\jpsi \, m_\jpsi),
\label{Optimiazed:NRQCD:factorization:indirect}
\eeq
where $e_c e=2/3e$ denotes the electric charge of the $c$ quark.
$f_\jpsi$ denotes the $J/\psi$ decay constant, which is defined by
\beq
  \langle J/\psi|\bar{c}\gamma^\mu c|0\rangle=-f_{J/\psi} m_{J/\psi}
\varepsilon_{J/\psi}^{*\mu}.
\eeq

It is worth stressing that, being a nonperturbative parameter, $f_{J/\psi}$ is intimately linked with the LO NRQCD matrix element appearing in \eqref{NRQCD:factorization:direct}.
As a matter of fact, NRQCD factorization implies the $J/\psi$ decay constant to be further factorized into the LO NRQCD matrix element multiplied with
perturbative corrections:
\beq
 f_\jpsi =\sqrt{\dfrac{2}{m_{J/\psi}}} \langle \jpsi(\boldsymbol\epsilon) \vert \psi^\dagger \boldsymbol{\sigma}\cdot\boldsymbol{\epsilon}\chi(\mu_\Lambda) \vert 0 \rangle
 \left[1- {2C_F\alpha_s\over \pi} + {\cal O}(\alpha_s^2)\right]+ {\cal O}(v^2).
\eeq
Thus far the three-loop QCD corrections to $f_\jpsi$ has been available.
It is unfortunate that radiative QCD correction at each perturbative order appears to be substantially negative,
and the perturbative convergence for this observable might be rightfully questioned.

It is important to mention that the $\jpsi$ decay constant is a physical observable, which can be directly determined from the precisely measured leptonic width of
$\jpsi$ through
\begin{align}
\Gamma\left(J/\psi\rightarrow l^+l^-\right)= {4\pi e_c^2\alpha^2(m_\jpsi) \over 3} {f^2_{J/\psi}\over m_{J/\psi}}.
\end{align}

In ascertaining the QCD radiative corrections to the NRQCD SDCs, one usually treats all the quark-level diagrams at a given order of
$\alpha_s$ on an equal footing.
However, the traditional NRQCD approach has some drawbacks to our process.
When considering the next-to-leading order QCD correction in the indirect channel,
those two-loop diagrams with the gluon attached onto the $t$ quark and those with the gluon attached to the $c$ quark
(the rightmost diagram in Fig.~\ref{Feyn:Diagrams:Indir}) have very different origin.
While the former type of QCD correction is associated with a hard scale ($\sim m_H$), the latter is responsible for the NLO QCD correction to
the $J/\psi$ decay constant, where the typical gluon virtuality is much lower ($\sim m_\jpsi$).
It should be noted that these two types of diagrams are separately QCD and QED gauge invariant.
In what follows, we would like to adopt an optimized NRQCD approach, by treating these two types of diagrams differently.
In our analysis of the indirect channel, we discard all those two-loop diagrams with the gluon attached to the $c$ quark and only consider those diagrams
with gluon exchanged between the $t$ quark. Such a partial selection of diagrams is compensated by the procedure that we choose the
nonperturbative parameter that characterizes the formation of $\jpsi$ to be $f_\jpsi$, which can be precisely determined from experiment,
rather than the LO NRQCD matrix element, which is subject to certain theoretical uncertainty.
This strategy amounts to conducting a resummation of all the diagrams responsible for the QCD corrections to the $J/\psi$ decay constant to all orders in $\alpha_s$ and $v$.
Actually, in some preceding work on $H\to \jpsi+\gamma$~\cite{Bodwin:2013gca,Bodwin:2016edd}, this optimized NRQCD treatment
has already been tacitly employed~\footnote{Note that this optimized NRQCD approach has also been fruitfully applied
to the $e^+e^-\to \gamma^*\gamma^*\to J/\psi J/\psi$ process~\cite{Bodwin:2006yd,Sang:2023liy},
which results in a satisfactory perturbative convergence behavior,
as opposed to the predictions from traditional NRQCD factorization approach that suffer from substantial negative radiative corrections~\cite{Gong:2008ce}.}.

We remark that the coefficient ${\mathcal{F}}_\text{indir}$ introduced in \eqref{Optimiazed:NRQCD:factorization:indirect} is only
sensitive to the subprocess $H\gamma^*\gamma$, which has nothing to do with the hadronization of the charm quarks into $\jpsi$.
It can be split into the LO (one-loop) and NLO (two-loop) pieces:
\beq
{\mathcal{F}}_\text{indir}
={\mathcal{F}}_\text{indir}^{(0)}+{\mathcal{F}}_\text{indir}^{(1)}+\cdots,
\label{form:factor:indirect:decomposition}
\eeq
with the superscript indicating the power of $\alpha_s$.

\section{Outline of the calculation}
\label{Outline:Calculation}

In this section we briefly describe the calculation of the two-loop QCD corrections to $H\to \jpsi+\gamma$ in the optimized NRQCD approach.
Some representative diagrams are shown in Figs.~\ref{Feyn:Diagrams:Dir} and \ref{Feyn:Diagrams:Indir}.

The lowest-order contribution to the $H\to\gamma^*\gamma$ subprocess proceeds at one loop order,
and we retain the physical $t/W$ masses when calculating the indirect amplitude.
The next-to-leading order (NLO) QCD correction only arises from those two-loop diagrams with top quark
dressed with gluon exchange.
The 12 two-loop diagrams and the corresponding amplitudes are generated with the aid of the package {\tt FeynArts}~\cite{Hahn:2000kx},
and the Dirac/Lorentz/color algebras are manipulated by utilizing the package {\tt FeynCalc}~\cite{Mertig:1990an,Shtabovenko:2016sxi,Shtabovenko:2020gxv}.
Throughout this work we utilize dimensional regularization to regularize both UV and IR divergences.
The expression of the LO contribution in the indirect channel is UV/IR finite, which can be readily obtained in a closed form.

When investigating the NLO QCD correction to the indirect amplitude, we adopt the heavy quark mass counterterm $\delta Z_m$
determined in the on-shell renormalization scheme in QCD.  Since the quark Yukawa coupling is proportional to the quark mass in SM,
the mass counterterm determined in QCD also simultaneously serves the counterterm for the Yukawa coupling constant.
In this work, the relevant counterterm Lagrangian in Yukawa sector entails $t$, $b$ and $c$ (the latter two will be relevant for the direct channel):
\beq
\mathcal{L}_{\text{Yuk}}= -\sum_{Q=t,b,c} Z_{m_Q} m_Q \overline{Q} Q\left(1+2^{1/4}\sqrt{G_{F}}H\right).
\label{Yukawa:counterterm}
\eeq

We use the package \texttt{Apart}~\cite{Feng:2012iq} and \texttt{Kira}~\cite{Klappert:2020nbg} to conduct the partial fraction
and integration by parts reduction for the amplitude. We end up with $35$ Master Integrals (MIs) in the indirect channel, which are subsequently evaluated
with the aid of the package {\tt AMFlow}~\cite{Liu:2017jxz,Liu:2020kpc,Liu:2022chg} with high numerical accuracy.
The renormalized form factor in the indirect channel is UV and IR finite.

Next we apply the standard NRQCD approach to the direct $J/\psi$ production channel.
The familiar perturbative matching procedure is invoked to determine the SDC, by replacing the physical $J/\psi$ by a fictitious one, {\it viz.},
a free $c\bar{c}$ pair carrying the quantum number ${}^3S_1^{(1)}$. Evaluating the on-shell quark amplitude for
$H\rightarrow c\bar{c}({}^3S_1^{(1)})+\gamma$ is facilitated with the aid of covariant projection technique.
Since the projected accuracy is ${\cal O}(v^0)$, the NRQCD SDC can be directly extracted according to the strategy of region~\cite{Beneke:1997zp},
by neglecting the relative momentum between $c$ and $\bar{c}$ pair prior to conducting the loop integration,
which amounts to retaining only the hard region contribution.

There are about $118$ two-loop diagrams for $H\rightarrow c\bar{c}({}^3S_1^{(1)})+\gamma$ in the direct channel.
After tensor reduction, we end up with 160 two-loop MIs, which are again calculated using {\tt AMFlow}~\cite{Liu:2017jxz,Liu:2020kpc,Liu:2022chg}
with high numerical accuracy~\footnote{It might be possible to work out the two-loop MIs encountered in the direct channel of
the $H\to \jpsi+\gamma$ process in the closed form.
In a similar work about the NNLO QCD corrections to $e^+e^-\to \gamma^*\to \eta_c+\gamma$~\cite{Chen:2017pyi},
the two-loop MIs have been calculated analytically. Unfortunately the resulting two-loop SDC entails complicated elliptical integrals
and looks rather cumbersome. For the phenomenological purpose, in this work we are content with presenting the two-loop
corrections in a purely numerical format, albeit with exquisitely high accuracy.}.
After implementing the $c$ quark wave function and mass on-shell renormalization~\cite{Broadhurst:1991fy, Melnikov:2000zc},
and renormalizing the strong coupling constant to one-loop order under $\overline{\rm MS}$ prescription,
as well as renormalizing the Yukawa coupling in line with \eqref{Yukawa:counterterm}, we end up the UV finite SDC associated with ${\cal F}^{(2)}_{\rm dir}$.
Nevertheless, at this stage, the two-loop form factor in direct channel still contains a single IR pole,
with the coefficient as exactly specified in \eqref{NRQCD:SDC:dir:through:two:loop}. As demanded by NRQCD factorization,
this IR pole can be factored into vacuum-to-$\jpsi$ NRQCD matrix element in accordance with the $\overline{\rm MS}$ prescription,
so that we end up with the UV/IR finite two-loop SDC, which then acquires an explicit $\ln \mu_\Lambda$ dependence.
Ultimately, we are able to identify the non-logarithmic coefficient function $f^{(2)}(\tau)$ in \eqref{NRQCD:SDC:dir:through:two:loop} with high numerical accuracy.

\section{NRQCD short-distance coefficients through two-loop order}
\label{SDCs:through:two:loop}

In this section, we report the results of the UV/IR finite SDCs in both direct and indirect channels through two-loop order in the
optimized NRQCD approach. While the one-loop expressions have been presented in closed form, we are content with providing the two-loop SDCs in
purely numerical format. In the numerical calculation, we choose $m_H=125.20$ GeV, $m_W=80.37$ GeV, $m_t=173$ GeV~\cite{ParticleDataGroup:2024cfk}, $m_c=1.5$ GeV and $m_b=4.7$ GeV.

\subsection{Direct $J/\psi$ production channel}

\begin{figure}
\includegraphics[width=.5\textwidth]{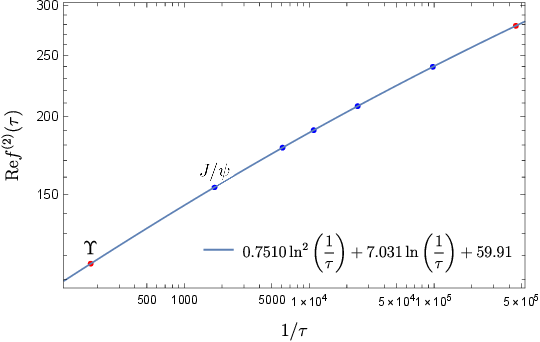}
\caption{The profile of the real part of $f^{(2)}(\tau)$. While the dots correspond to the actual two-loop calculation,
the curve represents the fit according to the parametrization implicated by asymptotic expansion (only the blue dots are used for the fit).
\label{Fig:f2:tau:and:asymptotic}}
\end{figure}

The LO SDC in the direct channel, as defined in \eqref{NRQCD:factorization:direct} and \eqref{form:factor:direct:decomposition} reads
\beq
{\mathcal{F}}^{(0)}_\text{dir}(\tau) = \dfrac{1}{1-\tau}.
\eeq

The respective coefficient function associated with the one-loop QCD correction, as introduced in \eqref{NRQCD:SDC:dir:one:loop},
can be worked out analytically:
\begin{align}
f^{(1)}(\tau) =& C_F \Bigg\{\dfrac{4-\tau(9-\tau)}{4(1-\tau)^2}
\left[\mathrm{Li}_2\left(\dfrac{\tau}{2-\tau}\right)+\dfrac{1}{2}\ln^2\left(\dfrac{\tau}{2-\tau}\right)+i\pi\ln\left(\dfrac{\tau}{2-\tau}\right)-\dfrac{\pi^2}{6}\right]\nn\\
  &-\dfrac{2(1-2\tau)}{(1-\tau)^2}\ln^2\left(\sqrt{\dfrac{1}{\tau}}-\sqrt{\dfrac{1}{\tau}-1}\right)
  -\dfrac{2\sqrt{1-\tau}(1-2\tau-\tau^2)+2i\pi(1-2\tau)}{(1-\tau)^2}\ln\left(\sqrt{\dfrac{1}{\tau}}-\sqrt{\dfrac{1}{\tau}-1}\right)
\nn\\
  &+\left[3-\dfrac{2}{1-\tau}-\dfrac{1}{2(2-\tau)}+\dfrac{1}{(2-\tau^2)}\right]\ln\left[\dfrac{\tau}{2(1-\tau)}\right]
  -\dfrac{3}{2}-\dfrac{1}{2(2-\tau)}
\nn\\
  &+\dfrac{1}{2}i\pi\left[6+2\sqrt{1-\tau}-\dfrac{1}{2-\tau}-\dfrac{4}{1-\tau}+\dfrac{2}
  {(2-\tau)^2}-\dfrac{8}{\sqrt{1-\tau}}+\dfrac{4}{(1-\tau)^{3/2}}\right]\Bigg\},
\end{align}
which is compatible with \cite{Shifman:1980dk}.

One of the major new results of this work is the two-loop QCD correction to the SDC in the direct channel.
The nontrivial QCD dynamics  is encapsulated in the coefficient function $f^{(2)}(\tau)$ in \eqref{NRQCD:SDC:dir:two:loop}.
As said before, here we are content with presenting this function in numerical format, whose
profile is depicted in Fig.~\ref{Fig:f2:tau:and:asymptotic}.
As is evident from Fig.~\ref{Fig:f2:tau:and:asymptotic}, in the limit $m_H\gg m_c$,
$f^{(2)}(\tau)$ is dominated by the double logarithm $\ln^2\tau$ asymptotically.
This logarithmic scaling is anticipated from the perspective of the collinear factorization and
ERBL evolution equation~\cite{Jia:2008ep}. In fact, we have explicitly verified that,
by combining the logarithm due to the $\overline{\rm MS}$ charm quark mass running  from $m_H$ to $m_c$
with the double collinear logarithm at two-loop order as predicted in \cite{Jia:2008ep}, we find perfect agreement
the fitted result, $0.7510\ln^2\tau$ in Fig.~\ref{Fig:f2:tau:and:asymptotic}.

Concretely for $H\to \jpsi+\gamma$ and $H\to\Upsilon+\gamma$, we have
\begin{subequations}
\bqa
&& f^{(2)}\left({4\times 1.5^2\over 125.2^2}\right)= -(179.954 +18.2043 i)+(11.2424 +0.276021  i) n_L,
\\
&& f^{(2)}\left({4\times 4.7^2\over 125.2^2}\right)=-(130.164 +14.8414 i)+(7.46541 +0.266299 i) n_L,
\eqa
\end{subequations}
with $n_L=3$ signifying the number of light quarks.
It turns out that those two-loop diagrams entailing light quark vacuum polarization only yield a small portion of contribution.

\begin{table}[htb]
\def\arraystretch{1.25}%
\begin{tabular}{|c|c|c|c|}
\hline
& ${\mathcal{F}}^{(0)}_\text{dir}$
& ${\mathcal{F}}^{(1)}_\text{dir}$
& ${\mathcal{F}}^{(2)}_\text{dir}$
\\\hline
$\jpsi$
& $1.00057$
& $-0.456508 + 0.115753 i$
& $-(0.28554 + 0.0288856 i) + (0.0178389 + 0.000437974 i) n_L$
\\\hline
$\Upsilon$
& $1.00567$
& $-0.372898 + 0.11637 i$
& $- (0.207589 + 0.0236694 i)+(0.011906 + 0.0004247 i) n_L $
\\\hline
\end{tabular}
\caption{Numerical values of ${\mathcal F}^{(i)}_\text{dir}$ (($i=0,1,2$)) in \eqref{form:factor:direct:decomposition} for $H\to \jpsi(\Upsilon)+\gamma$.
We have fixed $\mu_R=m_H/2$ and $\mu_\Lambda=m_{c(b)}$, and used {\tt RunDec}~\cite{Schmidt:2012az} to compute the QCD running coupling at
two-loop accuracy.
\label{Table:SDC:dir:channel}}
\end{table}

To gauge the relative importance of the QCD corrections at each perturbative order,
in Table~\ref{Table:SDC:dir:channel} we enumerate the numerical values of the SDCs ${\mathcal{F}}^{(i)}_\text{dir}$ ($i=0,1,2$) in \eqref{form:factor:direct:decomposition}.
Analogous to the one-loop corrections, the two-loop QCD corrections remain to be negative and significant in both processes.
The perturbative convergence of the SDC in the direct channel might look questionable.

\subsection{Indirect $J/\psi$ production channel}

We then proceed to the QCD correction in the indirect channel in \eqref{form:factor:indirect:decomposition},
where the corresponding Feynman diagrams are depicted in Fig.~\ref{Feyn:Diagrams:Indir}.

To our knowledge, the analytic expression of the one-loop amplitude for $H\rightarrow\gamma \gamma^*$ was first given in \cite{Bergstrom:1985hp}.
For the sake of completeness, we also present here the analytic expression of the LO contribution of ${\mathcal{F}}_{\text{indir}}$,
which can be divided into the $t$-loop induced piece and the $W$-loop induced piece~\footnote{Note that
our recipe of treating the indirect channel differs from the preceding work. For example, Bodwin {\it et al.}~\cite{Bodwin:2013gca}
combined the measured $H\to\gamma\gamma$ partial width with a phenomenological form factor, which is a function of the photon virtuality,
to estimate the effective $H\to\gamma\gamma^*$ vertex.}:
\begin{subequations}
 \begin{align}
  {\mathcal{F}}^{(0)}_{\text{indir},t} =& \dfrac{\sqrt[4]{2}e^2e_t^2N_c}{\pi^2}\dfrac{1}{32r_t^2(1-\tau)^2}\bigg\{\left[1-r_t(1-\tau)\right]\left[
  \arctan^2\left(\dfrac{2\sqrt{r_t \tau(1-r_t\tau)}}{1-2r_t\tau}\right)
  -\arctan^2\left(\dfrac{2\sqrt{r_t (1-r_t)}}{1-2r_t}\right)\right]
  \nn\\
  &+4\sqrt{r_t \tau(1-r_t\tau)}\arctan\left(\dfrac{2\sqrt{r_t \tau(1-r_t\tau)}}{1-2r_t\tau}\right)-4\sqrt{r_t (1-r_t)}\arctan\left(\dfrac{2\sqrt{r_t (1-r_t)}}{1-2r_t}\right)+4r_t(1-\tau)\bigg\},
  \\
  {\mathcal{F}}^{(0)}_{\text{indir},W} =& \dfrac{\sqrt[4]{2}e^2}{64r_W^2\pi^2(1-\tau)^2}
  \bigg\{\left[3-6r_W(1-\tau)+4r_W^2\tau(1-2\tau)\right]\left[\left(\pi-\arccos\left(2r_W -1\right)\right)^2-\arccos^2\left(1-2r_W \tau\right)\right]
\nn\\
  &+4\left[4r_W^2\tau-2r_W(1-\tau)-3\right]\Big[\sqrt{r_W\tau(1-r_W\tau)}\arccos\left(1-2r_W \tau\right)
\nn\\
&+\sqrt{r_W(1-r_W)}\tau\left(\arccos\left(2r_W -1\right)-\pi\right)+r_W(1-\tau)\Big] \bigg\}.
\end{align}
\label{H:gammagamma*:LO:expression}
\end{subequations}
where $e_t=2/3$, $r_t$ and $r_W$ are defined as
$$
r_t=\dfrac{m_H^2}{4m_t^2}, \qquad\qquad r_W=\dfrac{m_H^2}{4m_W^2}.
$$
Note that we have switched to a slightly different definition of $\tau$ in the indirect channel, $\tau\equiv m_\jpsi^2/m_H^2$.

Taking the $m_t,m_W\rightarrow\infty$ limit
, \eqref{H:gammagamma*:LO:expression} reduces to
\beq
\mathcal{F}^{(0)}_\text{indir,t}\rightarrow\dfrac{2^{1/4}\alpha e_t^2N_c}{ 3\pi}, \qquad\qquad
  \mathcal{F}^{(0)}_\text{indir,W}\rightarrow-\dfrac{7\alpha}{2^{7/4}\pi},
\eeq
which, reassuringly, recovers the well-known effective $H\gamma\gamma$ coupling
in the infinitely heavy $m_t$ and $m_W$ limit.

The NLO QCD correction to ${\mathcal{F}}_{\text{indir}}$ is computed numerically.
For $H\to \jpsi\gamma$ and $H\to\Upsilon\gamma$, we have
\begin{subequations}
\begin{align}
\mathcal{F}_\text{indir}^{(1)}\left(\dfrac{3.0969^2}{125.20^2}\right)=&-4.53641\times 10^{-5},
\\
\mathcal{F}_\text{indir}^{(1)}\left(\dfrac{9.4604^2}{125.20^2}\right)=&-4.53226\times 10^{-5}.
\end{align}
\label{SDC:indirect:QCD:correction}
\end{subequations}
In generating these numbers, we have resorted to the package {\tt PYTHIA}~\cite{Bierlich:2022pfr}
to infer the running QED coupling at one-loop accuracy and obtain $\alpha(m_H/2) =1/128.46$.

As opposed to the QCD corrections in the direct channel in Table~\ref{Table:SDC:dir:channel},
the NLO QCD corrections to the SDC in the indirect channel, \eqref{SDC:indirect:QCD:correction},
are real-valued and insensitive to the masses of the $c$ and $b$, and also appear to be rather insignificant in magnitude.

\section{Phenomenology}
\label{phenomenology}

In the phenomenological analysis, we take $G_F=1.1664\times 10^{-5}\,\mathrm{GeV}^{-2}$, $m_H=125.20$ GeV,
$m_W=80.37$ GeV, $m_t=173$ GeV, $m_{J/\psi}=3.0969\,\mathrm{GeV}$ and $m_\Upsilon=9.4604\,\mathrm{GeV}$~\cite{ParticleDataGroup:2024cfk}.
The central values of the quark masses are taken to be $m_c=1.5$ GeV and $m_b=4.7$ GeV. We use the package {\tt RunDec}~\cite{Schmidt:2012az} to compute
the QCD running coupling to two-loop accuracy. For the form factor in the direct channel, we take the QED coupling evaluated at the scale $m_H/2$,
$\alpha(m_H/2) =1/128.46$; for that in the indirect channel, we choose two of $\alpha$ evaluated at the scale of $m_H/2$, but take another
$\alpha$ evaluated at the vector quarkonium mass, {\it viz.},
$\alpha(m_\jpsi)= 1/132.77$ and $\alpha(m_\Upsilon)= 1/131.20$~\cite{Bierlich:2022pfr}.

In the optimized NRQCD approach, we need to specify the values of two interrelated nonperturbative parameters that encode the quarkonium formation information.
For the indirect channel, we take $f_\jpsi=403\,\mathrm{MeV}$ and $f_\Upsilon=685\,\mathrm{MeV}$, which are determined through the measured leptonic widths $\Gamma_{J/\psi}=5.53\;\mathrm{keV}$ and $\Gamma_{\Upsilon}=1.34\;\mathrm{keV}$, respectively.
For the direct channel, we estimate the value of the LO NRQCD matrix element from the quark potential model through
\beq
  \langle \jpsi(\boldsymbol\epsilon) \vert \psi^\dagger \boldsymbol{\sigma}\cdot\boldsymbol{\epsilon}\chi(\mu_\Lambda) \vert 0 \rangle \approx \sqrt{N_c\over 2\pi} R_\jpsi(0).
\label{estimate:NRQCD:LDME:potential:model}
\eeq
where $N_c=3$ is the number of colors, and $R_\jpsi(0)$ denotes the radial Shr\"odinger wave function of $\jpsi$ at the origin.
In this work we adopt the value of the radial wave functions at the origin evaluated from the Buchm\"uller-Tye (BT) potential model:
$R_\jpsi(0)=\sqrt{0.81}\;\mathrm{GeV}^{3/2}$, and $R_\Upsilon(0)=\sqrt{6.477}\;\mathrm{GeV}^{3/2}$~\cite{Eichten:1995ch}.
The matrix elements in \eqref{estimate:NRQCD:LDME:potential:model} are tacitly assumed to be evaluated at the default
NRQCD factorization scale $\mu_\Lambda=1\,\mathrm{GeV}$ for $J/\psi$, and $\mu_\Lambda=1.5\,\mathrm{GeV}$ for $\Upsilon$.

\begin{table}[htb]
\def\arraystretch{1.25}%
\begin{tabular}{|c|c|ccc|}
\hline
  \multicolumn{2}{|c|}{}
  &
 LO & NLO & NNLO \\
\hline
\multirow{2}{*}{$\jpsi$}
&$F_\text{dir}(10^{-8}\mathrm{GeV}^{-1})$
&$8.23456$
&$4.47758 + 0.952625 i$
& $(0.146811 + 0.00360445 i)+(2.26311 + 0.714902 i) n_L$
\\\cline{2-5}
&$F_\text{indir}(10^{-8}\mathrm{GeV}^{-1})$
& $-43.5765$
&$-43.9897$
& --
\\\hline
\multirow{2}{*}{$\Upsilon$}
&$F_\text{dir}(10^{-8}\mathrm{GeV}^{-1})$
& $-20.714$
&$-13.0333 - 2.39691 i$
& $ - (9.71754 + 1.90938 i)-(0.245231 + 0.00874764 i) n_L$
\\\cline{2-5}
&$F_\text{indir}(10^{-8}\mathrm{GeV}^{-1})$
& $12.1712$
&$12.2868$
& --
\\\hline
\end{tabular}

\caption{The form factors in direct and indirect channels at various perturbative accuracy for $H\rightarrow \jpsi(\Upsilon)+\gamma$.
\label{Table:Form:factor:dir:Indir}}
\end{table}

To assess the relative importance of the direct and indirect channels as well as the impact of QCD radiative corrections for $H\rightarrow \jpsi(\Upsilon)+\gamma$,
we enumerate in Table~\ref{Table:Form:factor:dir:Indir}  the values of $F_\text{dir}$ and $F_\text{indir}$ in \eqref{form:factor:divide:two:parts} at various perturbative accuracy.
In line with \eqref{form:factor:direct:decomposition} and \eqref{form:factor:indirect:decomposition}, the term
``NLO" indicates that $F_\text{NLO}\equiv F^{(0)}+F^{(1)}$ and ``NNLO" indicates that $F_\text{NNLO}\equiv F^{(0)}+F^{(1)}+F^{(2)}$.
It is clear that the direct and indirect amplitudes interfere destructively in both processes.
For $H\to J/\psi+\gamma$ process, the contribution from direct channel is overwhelmed by that from the indirect channel.
Therefore, despite significant negative ${\cal O}(\alpha_s^{1,2})$ corrections present in the direct channel,
the partial width is largely governed by the indirect channel contribution, and also by the interference between direct and indirect amplitudes.
The NLO QCD correction in the indirect channel turns out to be always insignificant.
Compared with $H\to J/\psi+\gamma$, the form factors in both direct and indirect channels for $H\to \Upsilon+\gamma$
bear comparable magnitudes, yet with opposite sign.
The NNLO prediction to the $F_\text{dir}$ is only slightly smaller than $F_\text{indir}$ in absolute magnitude, so that
the severe destructive interference renders the net form factor much smaller than that in each individual channel.

\begin{table}[htb]
\def\arraystretch{1.25}%
\begin{tabular}{|c|c|ccc|c|c|c|c|}
\hline
  \multicolumn{2}{|c|}{\multirow{2}{*}{}}
  &\multicolumn{3}{c|}{This work}
  &\multirow{2}{*}{\tabincell{c}{Bodwin\,\\\textit{et al.}~\cite{Bodwin:2013gca}}}
  &\multirow{2}{*}{\tabincell{c}{Bodwin\,\\\textit{et al.}~\cite{Bodwin:2014bpa}}}
  &\multirow{2}{*}{\tabincell{c}{Bodwin\,\textit{et al.}\\~\cite{Bodwin:2016edd,Bodwin:2017wdu}}}
  &\multirow{2}{*}{\tabincell{c}{Brambilla\,\\\textit{et al.}~\cite{Brambilla:2019fmu}}}
  \\
  \cline{3-5}
  \multicolumn{2}{|c|}{} & LO & NLO & NNLO & & & & \\
\hline
\multirow{2}{*}{$\jpsi$}
&$\Gamma_\text{indir}(\mathrm{eV})$
& $14.80$
& $15.08_{-0.03}^{+0.17}{}^{+0}_{-0}$
& --
& $13.2$
& ${14.2}^*$
& $-$
& ${13.8}^*$
\\\cline{2-9}
&$\Gamma(\mathrm{eV})$
& $9.74^{+0}_{-0}{}_{-0.29}^{+0.31}$
& $12.18_{-0.26}^{+1.60}{}^{+0.24}_{-0.23}$
& $13.29_{-0.28}^{+1.23}{}^{+0.24}_{-0.23}$
& $10.0$
& $11.7$
& $12.2$
& $12.3$
\\\hline
\multirow{2}{*}{$\Upsilon$}
&$\Gamma_\text{indir}(\mathrm{eV})$
& $1.1371{}$
& $1.1588_{-0.0022}^{+0.0052}{}^{+0}_{-0}$
& --
& $1.02$
& ${1.11}^*$
& $-$
& ${1.08}^*$
\\\cline{2-9}
&$\Gamma(\mathrm{eV})$
& $0.5602^{+0}_{-0}{}^{+0.0601}_{-0.0581}$
& $0.0484_{-0.0000}^{+0.0289}{}^{+0.0068}_{-0.0050}$
& $0.0546_{-0.0127}^{+0.0304}{}^{+0.0102}_{-0.0080}$
& $0.0574$
& $0.0026$
& $0.0213$
& $0.0408$
\\\hline
\end{tabular}
\caption{Our predicted partial widths of $H\rightarrow \jpsi(\Upsilon)+\gamma$ in comparison with other work.
 The first uncertainty originates from varying $\mu_R$ from $2m_{\jpsi(\Upsilon)}$ to $m_H$,
 whereas the second uncertainty is estimated from sliding $m_c$ from $1.3$ to $1.7$ GeV and
 sliding $m_b$ from $4.5$ to $4.9$ GeV.
 The numbers with asterisk are evaluated manually according to the equations provided by the corresponding literature.
\label{Table:partial:width}}
\end{table}

Substituting the form factors tabulated in Table~\ref{Table:Form:factor:dir:Indir} into \eqref{partial:wdith:formula:H:to:jpsi:gamma},
we are ready to predict the partial widths of $H\rightarrow \jpsi(\Upsilon)+\gamma$.
In Table~\ref{Table:partial:width}, we present our predictions of the complete decay widths and those solely from the indirect channel
at various perturbative accuracy, juxtaposed with the other predictions available in literature.
The entry ``LO"  in $\Gamma$ is obtained by retaining only the LO results of the $F_\text{dir}$ and $F_\text{indir}$ in \eqref{partial:wdith:formula:H:to:jpsi:gamma}.
The entry ``NLO" is obtained by keeping $F_\text{dir}^{(0)}+F_\text{dir}^{(1)}+F_\text{indir}^{(0)}+F_\text{indir}^{(1)}$,
whereas ``NNLO" is generated by further including $F_\text{dir}^{(2)}$ in  \eqref{partial:wdith:formula:H:to:jpsi:gamma}.

For $H\to \jpsi+\gamma$, the most precise prediction to the partial width, which is marked by NNLO, is somewhat enhanced with respect to the NLO prediction,
but somewhat smaller than the predicted $\Gamma_{\rm indir}$ at NLO accuracy. This shows that the negative ${\cal O}(\alpha_s^2)$ correction in the direct channel
has noticeable effect, which somewhat dilutes the effect of destructive interference.

On the other hand, since the direct and indirect amplitudes bear comparable magnitudes yet opposite sign,
the $H\to \Upsilon+\gamma$ process is quite sensitive to the higher order QCD corrections in the direct channel, because it can pronouncedly
affect the degree of destructive interference. As can be seen in Table~\ref{Table:partial:width}, our NNLO prediction to the complete decay width of
$H\to \Upsilon+\gamma$ is about one order-of-magnitude smaller than the LO prediction, but is somewhat greater than the NLO prediction. 
The strongly suppressed partial width indicates that the prospect of observing the $H\to \Upsilon+\gamma$ channel in
future experiments is rather gloomy.

\begin{figure}[hbt]
 \includegraphics[width=0.43\textwidth]{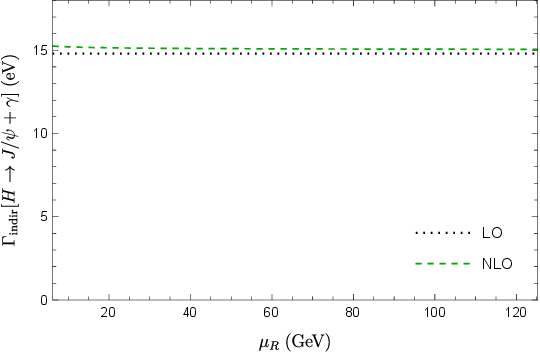}
 \includegraphics[width=0.43\textwidth]{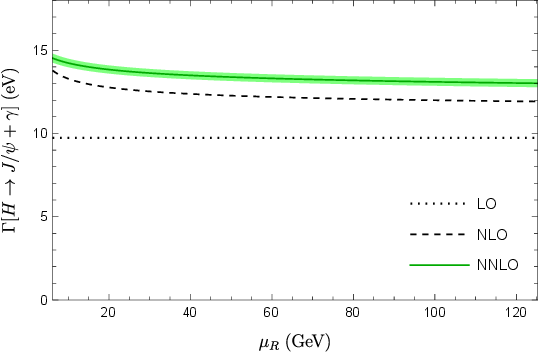}\\[1.25em]
 \includegraphics[width=0.43\textwidth]{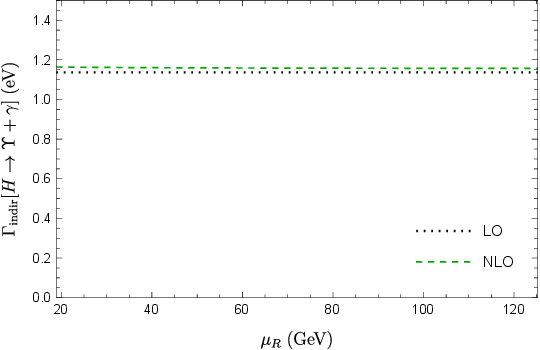}
 \includegraphics[width=0.43\textwidth]{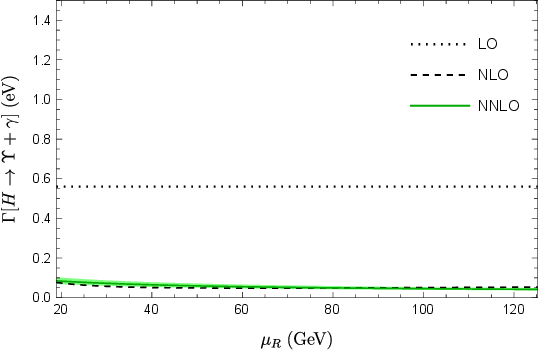}
 \caption{The partial width solely from the indirect channel and the complete partial width, as a function of $\mu_R$, with different levels of perturbative accuracy:
 $H\rightarrow\jpsi+\gamma$ (upper row), $H\rightarrow\Upsilon+\gamma$ (lower row).
 The green bands are obtained by varying $m_c$ from $1.3$ to $1.7$ GeV, and by varying $m_b$ from $4.5$ to $4.9$ GeV.
 \label{Plot:partial:widths:against:muR}}
\end{figure}

Supplementary to Table~\ref{Table:partial:width}, we also plot in Fig.~\ref{Plot:partial:widths:against:muR} the predicted partial widths of
$H\rightarrow\jpsi(\Upsilon)+\gamma$ against the renormalization scale. The contribution from the direct channel has much stronger $\mu_R$
dependence than that from indirect channel. Consequently, from the right panels of Fig.~\ref{Plot:partial:widths:against:muR}
we observe that the complete decay widths exhibits steeper $\mu_R$ dependence than the indirect decay width due to destructive interference.
Relative to the NLO prediction, the most refined NNLO predictions for $H\to\jpsi(\Upsilon)+\gamma$ do have a milder sensitively to $\mu_R$.

Finally we present our finest predictions to the branching fractions for $H\to\jpsi(\Upsilon)+\gamma$, based on the NNLO predictions of
the partial widths in Table~\ref{Table:partial:width}. To date the total width of Higgs boson measured at {\tt LHC} still bears large uncertainty, $\Gamma_H=3.2^{+2.8}_{-2.2}\,\mathrm{MeV}$~\cite{CMS:2019ekd}. To predict the branching fractions, we choose to use the much more precise prediction of
the Higgs full width, $\Gamma_H=4.07^{+4.0\%}_{-3.9\%}\,\mathrm{MeV}$~\cite{CMS:2019ekd,LHCHiggsCrossSectionWorkingGroup:2016ypw},
\begin{subequations}
\bqa
&& \mathcal{B}\left(H\rightarrow \jpsi+\gamma\right) = 3.27_{-0.07}^{+0.30}{}^{+0.06}_{-0.06}{}_{-0.13}^{+0.13}\times 10^{-6},
\\
&&  \mathcal{B}\left(H\rightarrow \Upsilon+\gamma\right) = 1.34_{-0.31}^{+0.75}{}^{+0.25}_{-0.20}{}^{+0.05}_{-0.05}\times 10^{-8}.
\eqa
\end{subequations}
The exceedingly tiny branching fraction of $H\rightarrow \Upsilon+\gamma$ make it unlikely to be observed in foreseeable future.
However, it is still worth looking for the clean $H\rightarrow \jpsi+\gamma$ signal at {\tt HL-LHC} and
future Higgs factories such as {\tt FCC} and {\tt CEPC}.

\section{Summary}
\label{summary}

In this work, we have calculated the two-loop QCD corrections to the Higgs boson radiative decay into $\jpsi$ and $\Upsilon$ using an optimized
NRQCD factorization approach, in which the direct and indirect channels are treated differently. The former process, with very clean experimental signature,
is potentially an attractive channel to extract the charm quark Yukawa coupling. 
It turns out that the NNLO prediction to the complete decay width is close to the LO prediction by solely including the indirect channel contribution,
and the negative ${\cal O}(\alpha_s^2)$ correction in the direct channel
has somewhat dilutes the effect of destructive interference.
For $H\to \Upsilon+\gamma$ process, since the direct and indirect amplitudes bear comparable magnitudes yet opposite sign,
the partial width becomes  quite sensitive to the higher order QCD corrections in the direct channel.
We present by far the finest predictions for the branching fractions of
Higgs boson radiative decay into vector quarkonia, $\mathcal{B}(H\rightarrow \jpsi+\gamma) = 3.27_{-0.07}^{+0.30}{}^{+0.06}_{-0.06}{}_{-0.13}^{+0.13}\times 10^{-6}$,
and $\mathcal{B}(H\rightarrow \Upsilon+\gamma)= 1.34_{-0.31}^{+0.75}{}^{+0.25}_{-0.20}{}^{+0.05}_{-0.05}\times 10^{-8}$.

\begin{acknowledgments}
We are indebted to Hai-Tao Li and De-Shan Yang for useful discussions. We are also grateful to Long-Bin Chen for participating in the early stage of this work.
This work is supported in part by
the National Natural Science Foundation of China under Grants No. 11925506.
The work of Z. M. is also supported in part by the National Natural Science Foundation of China No. 12347145, No. 12347105.
The work of J.-Y.~Z. is also supported in part by the US Department of Energy (DOE) Contract No.~DE-AC05-06OR23177, under which Jefferson Science Associates, LLC operates Jefferson Lab.
\end{acknowledgments}

\end{document}